\documentclass[a4paper,12pt]{article}
\usepackage{booktabs}
\usepackage{booktabs, makecell, longtable}
\usepackage{physics}
\usepackage{rotating}
\usepackage{makecell}
\usepackage[flushleft]{threeparttable}
\usepackage{amssymb}
\usepackage{graphicx}
\usepackage{amsmath}
\usepackage{epstopdf}
\usepackage[hang]{subfigure}
\usepackage{bm}
\usepackage[utf8x]{inputenc}
\usepackage{multicol, multirow}
\usepackage{url}
\usepackage[breaklinks,colorlinks,citecolor=blue,linkcolor=blue]{hyperref} 
\usepackage{authblk}
\urldef{\mailsa}\path|{silee}@sejong.ac.kr| 
\newcommand{\keywords}[1]{\par\addvspace\baselineskip
\newcommand{\RNum}[1]{\uppercase\expandafter{\romannumeral #1\relax}}
\noindent\keywordname\enspace\ignorespaces#1}
\newcommand{\RNum}[1]{\uppercase\expandafter{\romannumeral #1\relax}}
\newcolumntype{L}[1]{>{\raggedright\arraybackslash}p{#1}}
\newcolumntype{C}[1]{>{\centering\arraybackslash}p{#1}}
\newcolumntype{R}[1]{>{\raggedleft\arraybackslash}p{#1}}

\begin{document}

\title{Deeply Equal-Weighted Subset Portfolios}

\author{Sang Il Lee%
\thanks{Electronic address: \texttt{sangillee.fin@gmail.com}}}

\affil{DeepAllocation Technologies}

\maketitle

\begin{abstract}
The high sensitivity of optimized portfolios to estimation errors has prevented their practical application.
To mitigate this sensitivity, we propose a new portfolio model called a Deeply Equal-Weighted Subset Portfolio (DEWSP). DEWSP is a subset 
of top-$N$ ranked assets in an asset universe, the members of which
are selected based on the predicted returns from deep learning algorithms and are equally weighted.
Herein, we evaluate the performance of DEWSPs of different sizes $N$ in comparison with the performance of other types of portfolios such as optimized portfolios and historically equal-weighed subset portfolios (HEWSPs), which are subsets of top-$N$ ranked assets based on the historical mean returns. We found the following advantages of DEWSPs:
First, DEWSPs provides an improvement rate of 0.24$\%$ to 5.15$\%$ in terms of monthly Sharpe ratio compared to the benchmark, HEWSPs.
In addition, DEWSPs are built using a purely data-driven approach rather than relying on the efforts of experts. 
DEWSPs can also target the relative
risk and return to the baseline   
of the EWP of an asset universe by adjusting 
the size $N$.   
Finally, the DEWSP allocation mechanism is transparent and intuitive.
These advantages make DEWSP competitive in practice.

\end{abstract}

\section{Introduction}
Despite the significant success of deep learning,
its application to stock trading remains extremely challenging owing
to the volatile movements of stock prices, making it difficult to define the input values and understand how to apply the output values.
Machine learning models are built on a training set and are tested on a disjointed test set to prove their generalization capability,
and are commonly applied in various applications such as image processing, image recognition, speech recognition, and Internet searches.
However, this approach is limited when applied to the financial field owing to the time-evolving properties of the financial markets,
for example, structural breaks at occasional time points \cite{ludvigson2015uncertainty, stock2010modeling}, volatility clustering \cite{cont2007volatility}, and time-varying mean returns
 \cite{engle1987estimating}.
Furthermore, the time ordering of
financial data prevents the use of cross-validation
as a reliable estimate of the ensemble generalization error.
As a result, the performance of financial time-series models tends to be extremely sensitive to pre-specified periods, showing
the high power of in-sample (IS) prediction and the poor power of out-of-sample (OOS) prediction \cite{campbell2007predicting}.
This hampers the practical use of portfolio optimization techniques because an optimization is prone to the `garbage in, garbage out' phenomenon, in which biases occur in a portfolio selection unless predictions are adjusted suitably for an estimation error. To mitigate this problem, we built a new model called a Deeply Equal-Weight Subset Portfolios (DEWSPs) that combines deep learning techniques with an equal-weight strategy.
\\
\\
\noindent {\bf Portfolio theory}
The mean-variance portfolio (MVP) theory, pioneered by Markowitz (1952) \cite{markowitz1952portfolio}, has
long been recognized as the cornerstone of modern portfolio theory (MPT).
It provides a mathematical framework for determining a set of portfolios
with a maximized expected return per unit of risk; in addition,
the return and risk of a set are drawn as a line, called
an efficient frontier, on a risk-return plane.
However, despite the theoretical
advances in portfolio models including the MVP and its extensions,
their practical use remains problematic owing to
the difficulty in estimating 
reliable expected returns, which
critically affect the performance of
the portfolio \cite{jobson1980estimation}. 
For example, MVPs are not
necessarily well-diversified \cite{green1992will}, portfolio
optimizers are often ``error maximizers'' \cite{michaud2008efficient}, and
a mean–variance optimization can produce extreme or non-intuitive
weights for some of the assets in the portfolio \cite{black1991global,  
black1992global}.
 Many studies
have attempted to apply improved estimation procedures and mitigate the estimation error
problem. These include Bayesian methods \cite{jorion1986bayes, pastor2000portfolio}, shrinkage methods \cite{ledoit2000well} \cite{ledoit2003improved, wang2005shrinkage}, a
factor structure imposed on the returns \cite{mackinlay1999asset}, and the combination of 
a tangency portfolio, a risk-free rate, and a global minimum variance portfolio \cite{kan2007optimal}.
\\
\\
\noindent {\bf Equal-weight portfolio (EWP)}
There is a growing body of evidence showing that the use of simple rules of thumb is more successful than optimization. The most well-known example is the EWP, also called $1/N$ naive diversification, which is free of parameter uncertainty and has the following properties:
It never shorts any assets, it avoids a concentration, and upon a rebalancing of the dates, it sells high and buys low, thus exploiting a possible mean-reversion effect \cite{kritzman2010defense}.
The strength of an EWP is well known experimentally  \cite{jobson1980estimation, jorion1986bayes, demiguel2009optimal, duchin2009markowitz}.
DeMiguel et al.’s study \cite{demiguel2009optimal} is particularly convincing because
the authors evaluated 14 models on 7 empirical datasets. They
found that none of the 14 models consistently outperform a $1/N$ EWP. 
Tu and Zhou (2011) \cite{tu2011markowitz} showed 
the combination of an EWP and more sophisticated modes \cite{markowitz1952portfolio,jorion1986bayes,mackinlay1999asset,
kan2007optimal} is a way to improve performance.
The importance of the EW approach lies in its simplicity and widespread use.
In addition, Bernartzi and Thaler (2001) \cite{benartzi2001naive} 
demonstrated that EW diversification is ingrained in human behavior
by finding that a considerable fraction of participants equally distribute their contributions across the
available investment opportunities. 
This implies that investment decisions
tend to use intuition to choose a security and do not
necessarily rely on sophisticated formal techniques.
Investors can execute an EW strategy with a large universe but extremely low transaction costs using equally weighted exchange traded funds (ETFs),
for example, Direxion NASDAQ-100 Equal Weighted Index Shares, First Trust Dow 30 Equal Weight ETF, and
Goldman Sachs Equal Weight U.S. Large Cap Equity ETF. 
\\
\\
\noindent{\bf DEWSP} 
DEWSPs are constructed by incorporating deep learning techniques into an EW strategy. 
The building procedure consists of three steps:
First, the 1-month ahead return of assets is forecasted using deep learning algorithms. Second, assets are ranked in descending order based on the forecasts. Finally, subset portfolios are constructed with top-$N$ ranked assets that are equally weighted.
Our contribution is as follows:
\begin{itemize}
\item DEWSP is fully data-driven based on hyperparameter optimization. The entire process is automatic without
the views of human experts in building the models, which contributes to reduced
costs in terms of portfolio management. 
We also use public data on the prices and volume, which can be publicly obtained from various Web sites. Thus, DEWSPs are easily reproducible.
\item[•] 
DEWSPs show an increase in their risk and return
from the baseline of the EWP with a decrease in the number of assets.
This means that it is possible to control the aggressiveness of DEWSPs in terms of their risk-return tradeoff.
This mitigates difficulties in understanding
the black-box portfolio optimization and in 
tailoring the risk and return of ranked portfolios based on financial factors (e.g., based on size, value, and leverage).
\end{itemize} 
\noindent {\bf Related papers}
This study covers stock prediction using deep learning methods and ranked-portfolios.
Deep learning models are on the rise, showing impressive results in modeling the complex behavior of financial data. 
Examples include stock prediction based on long short-term memory (LSTM) networks \cite{fischer2018deep},
deep portfolios based on deep autoencoders \cite{heaton2017deep},
threshold-based portfolios using recurrent neural networks \cite{lee2018threshold},
deep factor models using deep feed-forward networks \cite{nakagawa2018deep}, a time-varying multi-factor model using LSTM networks \cite{nakagawa2019deep}, and an enhancing standard factor model using deep learning \cite{alberg2017improving}.

Ranked portfolios are widely used with
varying degrees of complexity, and
their basic premise is the same:
ranking stocks-based on 
factors such as 
their value, momentum, quality, size, low risk, and a combination of these factors, and then selecting a particular
proportion of the top-ranked stocks
to add to the portfolio.
These include
portfolios ranked in terms of size and book-to-market \cite{fama1995size}, portfolios ranked
on value and momentum factors \cite{asness2013value}, portfolios 
ranked on time-series momentum \cite{moskowitz2012time}, and
portfolios ranked on binary classification
using returns predicted through deep learning \cite{fischer2018deep}.
\\
\\
The remainder of this paper is organized as follows:
In Section \ref{sec:2}, we describe the data and preprocessing methods applied.
In Section \ref{sec:3}, we describe the experimental setting and implementation.
In Section \ref{sec:4}, we provide the experimental results and compare different portfolio models.
Finally, some concluding remarks are offered in section \ref{sec:5}.

\section{Data and preprocessing}
\label{sec:2}
\subsection{Universe}
Small portfolios are considered for an easier analysis, and are
important for several practical reasons \cite{maringer2006portfolio}:
First, it is difficult for small investors to acquire and continuously monitor a large portfolio. Second, large investors need to identify
a threshold where the cost exceeds the benefit of risk reduction from diversification.
Third, large portfolios amplify the estimation errors during the optimization process. To select a small but well-diversified universe,
we refer to the most commonly applied classification system, i.e., the Global Industry Classification Standard (GICS).
The asset universe consists of 
22 diversified stocks in Standard and Poor's 500 index
(S$\&$P 500) that belong to 11 different GICS sectors:
\begin{itemize}
\item 
{\bf Energy}: ExxonMobil (XOM) and Chevron (CVX),
{\bf Utilities}: Duke Energy (DUK) and Consolidated Edison (ED), 
{\bf Materials}: Sherwin-Williams (SHW) and DuPont (DD),
{\bf Industrials}: Boeing (BA) and Union Pacific (UNP), 
{\bf Consumer Discretionary}: Amazon (AMZN) and McDonald's (MCD) 
{\bf Consumer Staples}: Coca-Cola (KO) and Procter $\&$ Gamble (PG) 
{\bf Healthcare}: United Health Group (UNH) and Johnson $\&$ Johnson (JNJ)
{\bf Financials}: Berkshire Hathaway (BRK-B) and JPMorgan Chase (JPM) 
{\bf Information Technology Sector}: Apple (AAPL) and Microsoft (MSFT), 
{\bf Communication Services}: Facebook (FB) and Alphabet (GOOG), 
{\bf Real Estate}: American Tower (AMT) and Simon Property Group (SPG).
\end{itemize}

We use data from Yahoo Finance from January 1997 to October 2019, which is the common period of data availability. The monthly
stock dataset contains five attributes: open price, high price, low price, adjusted close price, and volume (OHLCV). The last of the daily OHLCV datasets per month is used as the raw dataset.
For each experiment, we split the
data into an in-sample ($70\%$) period and an out-of-sample ($30\%$) period.
The in-sample data are divided again into a training dataset (50$\%$) for developing the prediction models and a validation set (50$\%$) for evaluating its predictive ability.
\\
\\
\noindent {\bf Technical indicators} 
A technical analysis is a method for forecasting price movements using past prices and volume and includes a variety of forecasting techniques such as a chart
analysis, cycle analysis, and computerized technical trading systems. 

A technical analysis has a long history of widespread use by participants in
speculative markets 
\cite{smidt1965amateur,
billingsley1996benefits,
fung1997information,
menkhoff1997examining,
cheung2001currency,
gehring2003technical}, and
there is a large body of academic evidence
demonstrating
the usefulness of such analysis, including theoretical support 
\cite{brown1989technical} and empirical evidence 
\cite{lo2000foundations, blume1994market}, as well as the role of such analysis in out-of-sample equity premium predictability 
\cite{baetje2016equity,
rapach2010out,
neely2014forecasting}.
We used a full set of 14 technical indicators based on 3 types of popular technical strategies, i.e., the moving average crossover, momentum, and volume rules:
\begin{itemize}
 \setlength\itemsep{1em}
\item The time-series momentum indicator, MOM($m$), is the generation of a buy signal when the price is higher than the historical price. Its validation is supported by the observation that the ``trend'' effect persists for approximately 1 year and then partially reverses over a longer timeframe.  
Here, $\textrm{MOM}_{t}(m)$ at time $t$ is
defined as follows:
\begin{equation}
 \textrm{MOM}_{t}(m)=\begin{cases}
    1 \textrm{ (Buy signal) }, & \text{if} \quad P_{t} \geq P_{t-m}\\
    -1 \textrm{ (Sell signal) }, & \text{otherwise}.
  \end{cases}
\end{equation}
where $P_{t}$ is the index value at time $t$, and $m$ is the
look-back period. 
We use $m$ = 1, 3, 6, 9 and 12, which are respectively labeled as
$\textrm{MOM}_{t}$(1M), 
$\textrm{MOM}_{t}$(3M),
$\textrm{MOM}_{t}$(6M),
$\textrm{MOM}_{t}$(9M), and
$\textrm{MOM}_{t}$(12M).
	 
\item The moving average indicator, MA$(s,l)$, 
provides a signal for an upward or downward trend.
A buy signal is generated when the short-term moving average crosses above the long-term moving average because this represents the beginning of an upward trend. A sell signal is generated when the short-term moving average 
falls below the long-term moving average because this represents the beginning of a downward trend.

Let us define
a simple moving average of the index as follows:
\begin{equation}
\textrm{MA}_{j,t}^{P}=(1/j)\sum_{i=0}^{j-1}P_{t-m}  
\textrm{ for } j=s \textrm{ or }l,
\end{equation}	
where $s$ and $l$ are the look-back periods for short and long moving averages. 
The moving average indicator $\textrm{MA}_{t}(s,l)$
is then designed as follows:
\begin{equation}
 \textrm{MA}_{t}(s,l)=\begin{cases}
    1 \textrm{ (Buy signal) }, & \text{if} \quad \textrm{MA}_{s,t}^{P} \geq \textrm{MA}_{l,t}^{P}\\
    -1 \textrm{ (Sell signal) }, & \text{otherwise}.
  \end{cases}
  \end{equation}
	The six moving average indicators are constructed for $s$ =1, 2, and 3, and for
$l$ = 9 and 12, which are symbolized as 
	MA(1M-9M), MA(1M-12M), MA(2M-9M), MA(2M-12M), MA(3M-9M), and MA(3M-12M).
\item The volume indicator, VOL($s,l$), indicates a strong market
trend if the recent stock market volume and stock price increase.
Let us define the on-balance volume (OBV) as follows:
\begin{equation}
\textrm{OBV}_{t}=\sum_{k=1}^{t}VOL_{k}D_{k},
\end{equation}
where $VOL_{k}$ is a measure of the trading volume (i.e., number of shares traded) during period $k$, and $D_{k}$ is a binary variable: 
\begin{equation}
  D_{k}=\begin{cases}
    1, & \text{if} \quad P_{k}\geq P_{k-1}\\
    -1, & \text{otherwise}.
  \end{cases}
\end{equation}
The value of $\textrm{OBV}_{t}$ conceptionally measures both positive and negative 
volume based on the belief that changes in volume can predict a stock movement. The volume-based indicator is then defined as the difference between
the moving averages with an $s$-period and an $l$-period:
\begin{equation}
 \textrm{VOL}(s,l)=\begin{cases}
    1 \textrm{ (Buy signal) }, & \text{if} \quad \textrm{MA}_{s,t}^{\textrm{OBV}} \geq \textrm{MA}_{l,t}^{\textrm{OBV}}\\
    -1 \textrm{ (Sell signal) }, & \text{otherwise}.
  \end{cases}
\end{equation}
Here,
$
\textrm{MA}_{j,t}^{\textrm{OBV}}=(1/j)\sum_{i=0}^{j-1}\textrm{OBV}_{t-i}
$ is the moving average of $\textrm{OBV}_{t}$ for $j = s$ or $l$. 
The six moving average indicators are constructed for $s$ =1, 2, and 3 and 
for $l$ = 9 and 12, which are symbolized as follows:
VOL(1M-9M), VOL(1M-12M),
VOL(2M-9M), VOL(1M-12M), VOL(3M-9M) and VOL(3M-12M).
\end{itemize}

\section{Frameworks}
\label{sec:3}

\subsection{Portfolios}
For a comparative analysis, we also built three different types of portfolios, which are distinct in terms of their optimization or estimation process. 
All portfolios are built on the following assumptions: 
(1) all stocks are infinitely divisible; 
(2) there are no restrictions on the buying or selling of any selected portfolio; 
(3) there is no friction (e.g., transaction costs, taxation, commissions, or liquidity); and 
(4) it is possible to buy and sell stocks at the closing prices at any time $t$.
We adapt a periodic rebalancing strategy in which
the investor adjusts the weights in the investor’s portfolio
at the close price on the last business day of 
every month.
\\
\\
\noindent {\bf List of portfolios considered}:

\begin{itemize}
\item DEWSP:
This is a subset of portfolios that
consist of the top $N$-th ranked assets 
among all $N_{0}$ assets based on their expected returns.   

\item EW whole portfolio (EWWP): 
This is a traditional EWP of all assets $N_{0}$,
and can be viewed as a special case
of DEWSP when $N = N_{0}$.
Because there are no parameter estimations, 
it serves as the baseline  
for an evaluation of the risk and return of the DEWPs of different sizes.
\item Historically EW subset portfolios (HEWSPs): 
Like DEWSPs, HEWSPs are top-ranked subset portfolios, although their 
expected returns are estimated as a historical average over
the training and validation (HEWSP-TV) and historical average over
the validation (HEWSP-V).
This reveals the effect of the return prediction of the DEWSPs. 
\item Randomly EW subset portfolios (REWSPs): 
These are subsets of portfolios consisting of $N$ assets selected randomly, without the use of a ranking method.
A comparison between REWSPs and DEWSPs and HEWSPs reveals the effect of the estimated return prediction.
\item Maximum Sharpe ratio portfolios (MSRPs): 
These are complete portfolios that are maximized to achieve the highest
Sharpe ratio, and are mathematically defined as follows: 
\begin{eqnarray}
&\max_{\boldsymbol{w}_{t}}\boldsymbol{w}_{t}^{T} \boldsymbol{\mu}_{t}/\sqrt{\boldsymbol{w}^{T}_{t}\Sigma_{t} \boldsymbol{w}_{t}}
\quad \textrm{s.t. } \boldsymbol{w}_{t}^{T}\boldsymbol{1}=1, \textrm{ and }  w_{i,t} \geq 0, \textrm{ } \forall i,  
\end{eqnarray}
where $\boldsymbol{\mu}_{t}$ is a vector 
of $N_{0}$ predicted returns, $\boldsymbol{w}_{t}=(w_{1,t},\ldots,
w_{N_{0},t})^{T}$ is
a vector of portfolio weights,  
$\Sigma_{t}$ is a covariance matrix of the asset returns, 
$\boldsymbol{1}_{N}=(1,\ldots,1)^{T}$ is an $N$-dimensional
vector, and $\boldsymbol{w}_{t}^{T} \boldsymbol{\mu}_{t}$ and 
$\boldsymbol{w}_{t}^{T} \sum_{t} \boldsymbol{w}_{t}$ are the portfolio return and variance, respectively.
Because $\boldsymbol{\mu}_{t}$ and $\Sigma_{t}$ 
are unknown in practice, we replace them with
$\widehat{\boldsymbol{\mu}}_{t}$ from deep learning algorithms and $\widehat{\Sigma}_{t}$ from an in-sample dataset.
A comparison with DEWSPs reveals 
the effect of the estimation error on a portfolio optimization. 

\item Minimum variance portfolios (MVPs): 
These are complete portfolios optimized for the lowest
volatility, and solve the following constrained minimization problem:
\begin{equation}
\min_{\boldsymbol{w}} \boldsymbol{w}_{t}^{T}
\Sigma_{t}\boldsymbol{w}_{t} \quad \textrm{s.t. } \boldsymbol{w}_{t}^{T}\boldsymbol{1}=1, \textrm{ and }  w_{i,t} \geq 0, \textrm{ } \forall i.
\end{equation}
A comparison with DEWSPs reveals 
the effectiveness of optimization under the condition of no estimation errors.
\end{itemize}

\subsection{Prediction models}
\subsubsection{Training} 
A multilayer feedforward neural network (FFNN) was used in this study. We used Tree-structured Parzen Estimator (TPE) approach \cite{bergstra2011algorithms} for automated hyperparameter tuning and 
Table \ref{params} presents the list of hyperparameters and their values. 
Each optimization run was initialized with randomly selected points, after which it proceeded sequentially for a total of 50 function evaluations.
During one evaluation run, the FFNN was trained over an in-sample training data.
The mean squared error (MSE) is calculated on a validation set per function evaluation, early stopping was applied when there is no
improvement on the validation accuracy after 10 continuous epochs.

\subsubsection{Regularizer} 
We used two popular regularization methods, i.e., a dropout and batch normalization (BN). 
A dropout \cite{srivastava2014dropout} 
is a simple way to prevent co-adaptation among
hidden nodes of a deep feed-forward neural network by dropping out randomly selected hidden nodes.
In recent years, BN \cite{IoffeS2015batch} has replaced a dropout
in modern neural network architectures,
and uses the distribution of
the summed input to a specific neuron over a mini-batch of training cases to compute the
mean and variance, which are then used to normalize the summed input to that
neuron for each training case.
A dropout and BN layers were employed for all hidden layers.
\\
\begin{table}[h]
\centering
\caption{List of hyperparameters and range of each hyperparameter.}
\label{table:meanerrorbaseline}
\begin{tabular}{lll}
\toprule
  Hyperparamter &\quad \quad \quad & Considered values/functions \\
\midrule
    Number of Hidden Layers && \{2, 3\} \\
    Number of Hidden Units && \{2, 4, 8, 16\} \\
     \makecell[l]{ Standard deviation} &&\{0.025,0.05,0.075\}\\
    Dropout  && \{0.25, 0.5, 0.75\} \\
    Batch Size && \{28, 64, 128\} \\
    Optimizer && \{RMSProp, ADAM, SGD (no momentum)\} \\
    Activation Function&& Hidden layer: \{tanh, ReLU, sigmoid\}, Output layer: Linear \\
    Learning Rate && \{0.001\} \\
    Number of Epochs && \{100\} \\ 
\bottomrule
\end{tabular}
\parbox{\textwidth}{\small%
\vspace{1eX} 
{\bf Number of layers}: number of layers of a neural network.
{\bf Number of hidden units}: number of units in the hidden layers
of a neural network.
{\bf Standard deviation}: standard deviation of a random normal initializer. 
{\bf Dropout}: dropout rates. 
{\bf Bath size}: number of samples per 
batch.  
{\bf Activation}: sigmoid function $\sigma(z)=1/(1+e^{-z})$, hyperbolic 
tangent function $\textrm{tanh}(z)=(e^{z}-e^{-z})/(e^{z}-e^{-z})$,
and rectified linear unit (ReLU) function $\textrm{ReLU}(z)=\textrm{max}(0,z)$.
{\bf Learning Rate}: learning rate of the back-propagation algorithm.
{\bf The Number of Epochs}: number of iterations over all training data.
{\bf Optimizer}: stochastic gradient descent (SGD) \cite{kingma2014adam}, RMSProp \cite{tieleman2012lecture}, and ADAM \cite{kingma2014adam}}
 \label{params}%
\end{table}

\subsection{Evaluation metrics}
\noindent {\bf Average percent change (APC)}
The APC measures 
the rate of change in a DEWSP return and the volatility 
as size $N$ increases from $N = 1$ to $N = N_{0}$ 
to see the rate of change from the baseline
of $N = N_{0}$ to $N = 1$,
and is defined as follows:
\begin{equation}
\textrm{APC}_{x}=\frac{1}{N_{0}-1}\sum_{N=1}^{N_{0}-1}\frac{x^{N}-x^{N+1}}
{x^{N+1}},
\end{equation}
where $x$ is $r_{t}$ or $\sigma_{t}$.

\noindent {\bf Average Sharpe ratio improvement rate (ASRIR)} 
ASRIR measures the relative improvement of the DEWSPs as compared to the HEWSP benchmark in terms of the Sharpe ratio (SR), and is defined as follows: 
\begin{equation}
\textrm{ASRIR}=\frac{1}{N_{0}}\sum_{N=1}^{N_{0}}\frac{x_{\textrm{DEWSP}}^{N}-x_{\textrm{HEWSP-TV/T}}^{N}}
{x_{\textrm{HEWSP-TV/T}}^{N}},
\end{equation}
where $x$ is the SR of DEWSPs and HEWSPs of the same size $N$.

\section{Experiments and Results}
\label{sec:4}
We examined the portfolio performance over both IS and OOS periods for three different universes: a total of 22 stocks (Exp. I), with 11 stocks consisting of the first stocks of each sector on the list (Exp. II), and the other 11 stocks (Exp. III). The following observation was made based on the empirical simulation results.

\subsection{In-sample performance}
\begin{figure}[t]
\centering
     \scalebox{0.555}
     {
	\includegraphics{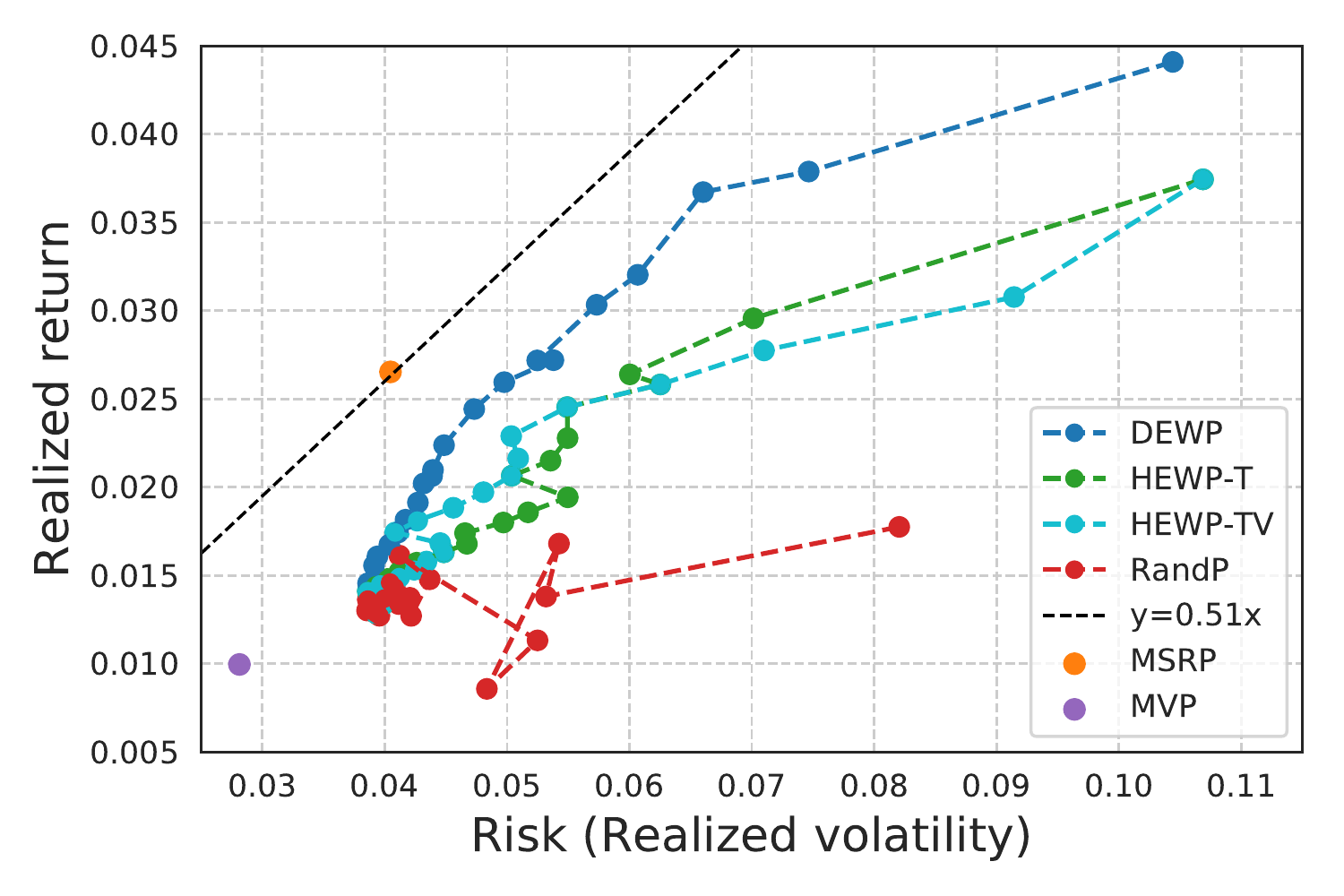}
	\includegraphics{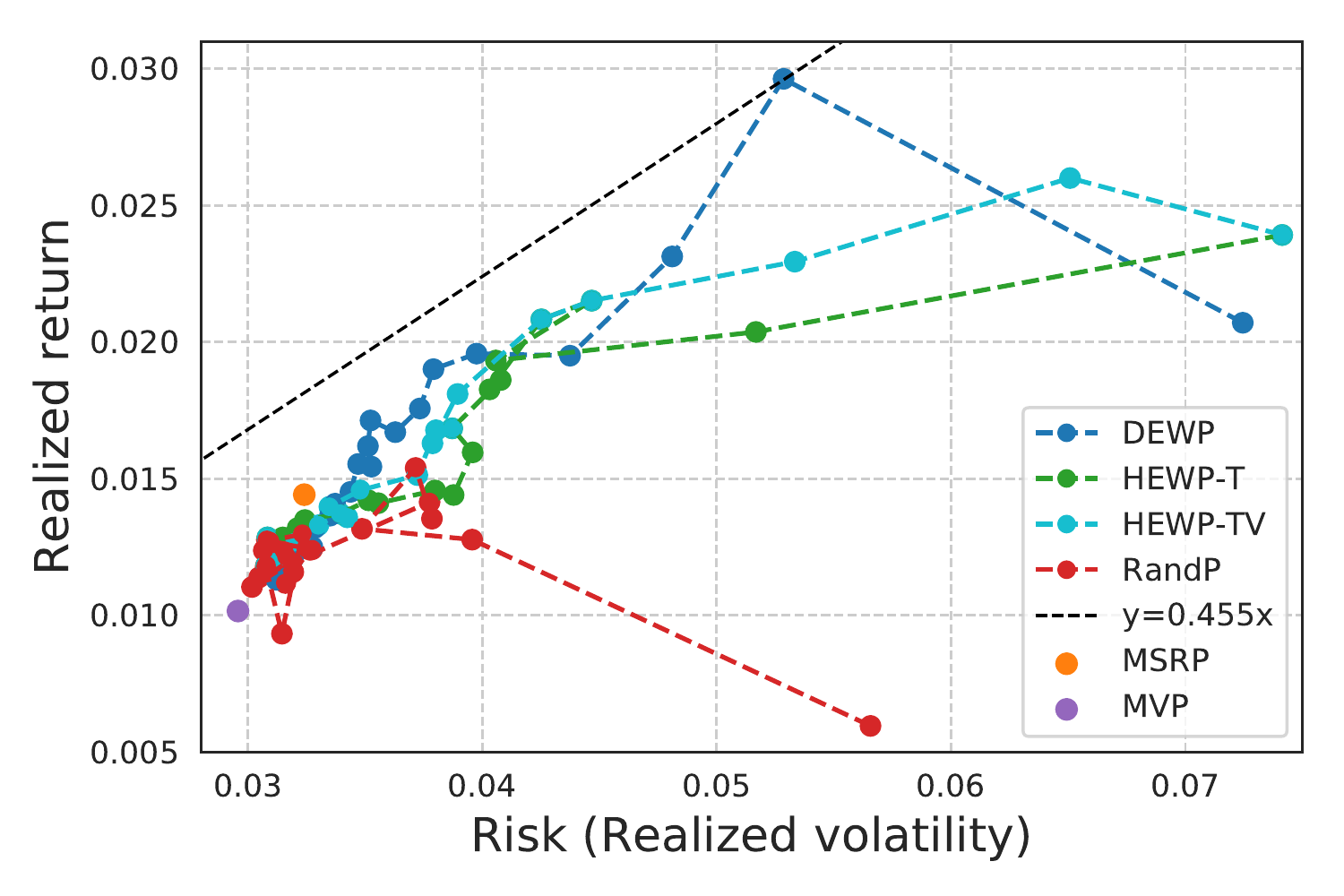}
}

\caption{
Realized risk vs. return of six different types of portfolios for the in-sample (left) and out-of-sample (right) experiments. The dotted lines specify the maximum SR estimate.}
\label{R_V_Frontier}
\end{figure}

\begin{itemize}
\item The left side of Figure \ref{R_V_Frontier} graphically shows the realized risk and return points of the portfolios on the risk-return plane. 
Each color represents a different type of portfolio, and different points with the same color represent 
different sizes. 
A comparison of DEWSPs and HEWSPs
with the REWSPs of a (seemingly) random pattern
indicates that the prediction-based ranking assets 
can be used to construct portfolios 
with increasing return and volatility
as $N$ decreases.
In Table \ref{APC1_val}, $\textrm{APC}_{r}$s and $\textrm{APC}_{\sigma}$s indicate quantitative measurements of the increase over Exp. I, II, and III, and $\textrm{APC}_{r}/\textrm{APC}_{\sigma}$
shows the degree of trade-off between the return and risk.  

\begin{table}[htbp]
  \centering
  \caption{Performance evaluation results of DEWSPs over in-sample period.}
    \begin{tabular}{lrrr}
           \toprule
  \multicolumn{1}{l}{Metrics ($\%$)}   & \multicolumn{1}{l}{Exp. I} & \multicolumn{1}{l}{Exp. II} & \multicolumn{1}{l}{Exp. III} \\
          \hline
 $\textrm{APC}_{r}$   & 6.11 &10.06 &11.23 \\
 $\textrm{APC}_{\sigma}$   & 5.07 & 8.86 &10.39\\
  $\textrm{APC}_r/\textrm{APC}_{\sigma}$   & 1.20 & 1.13 & 1.08\\ 
  ASRIR (w.r.t. HEWSP-TV) & 21.15 & 27.04&13.09\\ 
    ASRIR (w.r.t. HEWSP-T) & 19.12 & 21.93&12.24\\ 
     \bottomrule
    \end{tabular}%
  \label{APC1_val}%
\end{table}%

\item We also found ASRIRs of $21.15$, $27.04$, and $13.09\%$ for Exp. I, II, and III, respectively, indicating the superiority of DEWSP during the IS period.
\item MVP, as expected, achieves
the least volatility of $0.99$, and MSRP achieves the highest Sharpe ratio of 0.65 ($\mu$ = 0.026 and $\sigma$ = 0.040), which 
outperform those of the DEWSPs. This suggests that the prediction quality over the IS period is sufficient, allowing a benefit from the optimization process. 
\end{itemize}

\subsection{Out-of-sample performance}
\begin{itemize}
\item The DEWSPs are built using 1-month ahead predicted returns 
from the trained model, and the HEWSPs are 
built using the average historical return over
the in-sample period. 
\item 
The computation results are summarized on the right side of Figure \ref{R_V_Frontier} 
and in Table \ref{APC1_test}. 
As with the IS experiment, the return and volatility of the DEWSPs and HEWSPs still show an increasing pattern with the positive APC values. 
This allows us to tailor the portfolio's return and risk for investment purposes. 

\begin{table}[htbp]
  \centering
  \caption{Performance evaluation results of DEWSPs over the out-of-sample period.}
    \begin{tabular}{lrrr}
           \toprule
  \multicolumn{1}{l}{Metrics ($\%$)}   & \multicolumn{1}{l}{Exp. I} & \multicolumn{1}{l}{Exp. II} & \multicolumn{1}{l}{Exp. III} \\
          \hline
 $\textrm{APC}_{r}$  & 3.24 & 6.74 & 7.45 \\
 $\textrm{APC}_{\sigma}$    & 4.42 & 6.64 & 8.86\\
  $\textrm{APC}_r/\textrm{APC}_{\sigma}$   & 0.73 & 1.01 & 0.84\\ 
    ASRIR (w.r.t. HEWSP-TV) & 3.91 & 5.15 &0.24\\ 
    ASRIR (w.r.t. HEWSP-T)& 3.30 & 3.91 &0.92\\ 
   \bottomrule
    \end{tabular}%
  \label{APC1_test}%
\end{table}%

\item The ASRIRs ranged from 0.24 to 5.15$\%$ indicate that the DEWSPs outperform the historical models in terms of the monthly SR. The values are small compared to 
those of the in-sample ASRIRs, but indicate 
promising results. First, we can beat 
the HEWSP benchmark, and second, we
can tailor the return and volatility of the portfolios 
relative to the baseline of the EWWP. 

\item Although the MVP without a parameter estimation still gives the least volatility at $0.295\%$,
the MSRP has a monthly SR of $0.44\%$ ($\mu = 0.014$ and $\sigma = 0.032$), which is lower than that of the DEWSPs and HEWSPs, indicating that the prediction quality is insufficient for the purposes of portfolio optimization.
\end{itemize}
  
\section{Conclusion}
\label{sec:5}
Despite the significant success of machine learning in numerous fields,
stock prediction is still severely limited owing to its seasonal, non-stationary, and unpredictable nature. Consequently, portfolio models are inevitably exposed to the risk of estimation errors, 
which hinders their performance. 

To cope with such risk, we have proposed a new DEWSP model by incorporating deep-learning-based predictions into the framework of the EW strategy. We empirically demonstrated that DEWSPs 
can be used to target the levels of portfolio return and risk 
relative to the baseline of the EWWPs by adjusting the number of assets, and that its mechanism is clear in terms of the risk-return trade-off. We also showed that DEWSPs
are superior to HEWSPs in terms of the SR and  
that the mean-variance optimization amplifies the estimation error dramatically, which results in a substantially worse Sharpe ratio.
To summarize, DEWSPs are attractive from an implementation perspective, i.e., the use of public stock data, a transparent mechanism based on a risk-return trade-off, automatic hyperparameter optimization, 
the existence of a baseline of the EWP and a benchmark of the HEWSP, the capability of building portfolios using small numbers of assets (with expandability to large assets), and a simple incorporation of deep learning algorithms into the portfolio scheme. 

\bibliographystyle{unsrt}
\bibliography{BibFile}

\end{document}